**A near infrared emission feature from visible fluorescence tails and its correlation with singlet oxygen**

*Bjoern F. Hill[1], Jiaqi Li[1], Norman Labedzki[1], Christina Derichsweiler[1,2], Janus A. C. Wartmann[1], Krisztian Neutsch[1], Luise Erpenbeck[3], Sebastian Kruss,[1,2] ***

[1] Department of Chemistry and Biochemistry, Ruhr Universität Bochum, 44801 Bochum, Germany

[2] Fraunhofer Institute for Microelectronic Circuits and Systems, 47057 Duisburg, Germany

[3] Department of Dermatology, University Hospital Münster, 48149 Münster, Germany.

**Abstract**

The near infrared (NIR, >800 nm) or short wave infrared (SWIR) range of the electromagnetic spectrum overlays with the tissue transparency window and the telecommunication window, which makes it important for fluorescence-based imaging and photonics in general. Additionally, there are NIR absorption features related to vibrational overtone or combination modes and autofluorescence in photosynthetic organisms such as plants.

Here, we report a persistent NIR tail and feature at 923 nm that is observed during the photochemical generation of singlet oxygen using various photosensitizers. This feature initially appears to originate from singlet oxygen as its energy fits to known higher order optical and vibrational transitions, and it occurs simultaneously with the characteristic singlet oxygen phosphorescence at 1275 nm. To determine its origin, a systematic investigation is carried out using different photosensitizers (Rose Bengal, a ruthenium complex, Methylene Blue, chlorophyll), solvents, excitation sources and optical setups. The emission is present under all photochemical conditions, but it is not observed for chemically generated singlet oxygen. Its intensity correlates with photosensitizer concentration (1 µM – 1 mM), excitation intensity and the visible fluorescence intensity, but does not require the presence of oxygen as the 1275 nm phosphorescence of singlet oxygen.

A simulation shows that this feature arises from the NIR-tail of (visible) photosensitizer emission and the non-linear increase of detector sensitivity in the range between 900 nm - 950 nm. At the same time this feature and the whole NIR tail correlates with photochemically generated singlet oxygen concentration under certain conditions. Thus, careful interpretation of NIR emission is crucial but it also contains valuable information.

**Introduction**

Fluorescent materials play an important role in a wide range of applications, particularly in biosensing. They are used for molecular imaging, targeted diagnostics, and as optical labels[1–3]. Near infrared (NIR)/short wave infrared (SWIR) fluorescence offers major advantages. The NIR II window above 1000 nm is particularly well-suited for the study of biological systems (tissue transparency window)[1,4]. Many biological tissues and fluids exhibit strong autofluorescence in the visible spectral range under ultraviolet or visible light excitation. In the NIR range, this background autofluorescence is drastically reduced, improving the specificity and dynamic range of fluorescence-based labeling sensing[3,5]. Furthermore, absorption and scattering by biological tissue is strongly reduced compared to the visible spectral range[3,6]. This combination allows NIR radiation to achieve higher tissue penetration up to several millimeters, an improved signal-to-noise ratio and minimized phototoxicity[1]. These properties offer real-time, high-contrast visualization of biological processes making NIR fluorophores particularly attractive for *in vivo* imaging, long-term cell tracking, and non-invasive diagnostic techniques[1,5]. A broad range of fluorescent materials and molecules have been developed and applied for NIR-bioimaging, including organic dyes and fluorescent nanomaterials. Organic NIR dyes such as cyanines [7–9] and modified boron-dipyrromethenes[8,10] have been successfully applied for biological studies due to their tunable spectral properties and straightforward synthesis but they can suffer from bleaching, low quantum yield or biocompatibility.

In addition, many NIR-fluorescent nanomaterials have been explored. For example, semiconducting single-walled carbon nanotubes (SWCNTs) emit in the NIR and offer advantages because they do not bleach or blink[11,12]. They can be chemically functionalized to form highly selective biosensors[13–17]. More recently, 2D materials obtained by exfoliation have been investigated. These include Egyptian blue, a silicate with a high photostability and biocompatibility making it a promising candidate for use in NIR biosensors and bioimaging[18–20]. Other NIR-emitting nanomaterials include quantum dots[21] or doped nanoparticles[22].

These NIR fluorophores have been used for biosensing in complex biological environments. *In vitro* and *vivo* imaging benefit greatly from NIR fluorophores, including clinical applications. NIR fluorescent sensing has also been carried out to detect reactive oxygen species[23,24], small molecules[25,26], neurotransmitters[27–30] or pathogens[31]. The quantum yield of NIR fluorophores is in many cases lower than that of dyes in the visible range. For Egyptian Blue, values of 0.02 – 0.3 have been measured[32]. Both, common organic NIR dyes and nanoparticles such as quantum dots exhibit quantum yields from 0.01 to 0.15 [3,10,21]. However, the advantages of drastically reduced background fluorescence, absorption, and scattering outweigh the comparatively low quantum yields. Singlet oxygen ($^1O_2$) is a reactive oxygen species that plays a crucial role in numerous chemical and biological processes, including photodynamic therapy, oxidative stress and cellular signaling[33,34].

Molecular oxygen is in its ground state in a triplet state, with two unpaired electrons in antibonding π-orbitals. Excitation to the singlet state results in paired electrons, making it highly reactive but also short-lived[35,36]. Due to its short lifetime and reactivity, direct detection of $^1O_2$ is analytically challenging[36]. Other reactive oxygen species such as $H_2O_2$ can be effectively detected with fluorescent nanosensors, whereas the detection of $^1O_2$ with NIR sensors has been less investigated[37,38]. In addition to a luminescence in the red spectral range at around 760 nm, it also exhibits a weak luminescence in the NIR around 1275 nm. This NIR emission is used to detect $^1O_2$, as it serves as a direct sign of singlet oxygen formation and has been used in a number of spectroscopic studies and applications[35,36].

Detection of NIR fluorescence requires specialized optical instrumentation. The sensitivity of silicon-based detectors for visible light imaging or spectroscopy rapidly decreases in the NIR range and is limited to wavelengths below ~1100 nm[39–41]. In contrast, indium gallium arsenide (InGaAs) cameras and spectrometers enable sensitive detection in the NIR and NIR-II ranges, typically covering 900 – 1700 nm[39]. In addition to the detectors, further optics such as long-pass or band-pass filters, which are often based on dielectric coatings, as well as lenses and mirrors must be adapted to the NIR range. Proper configuration of these components is critical for reliable signal acquisition. However, the low signal intensities in the NIR also increases the sensitivity to optical artifacts. Overcoming all these challenges is crucial to ensure high performance NIR (bio)photonics.

Here, we report a NIR emission feature when $^1O_2$ is photochemically generated. It first seems to fit to an electronic/vibrational transition of $^1O_2$. We then study its correlation to $^1O_2$ concentration and propose a mechanism, which is important for NIR imaging or spectroscopy studies in all systems that contain photosensitizers from organic dyes to plants.

**Results and discussion**

**A generic NIR tail and feature**

To investigate the potential of NIR-fluorescent nanomaterials for the detection of singlet oxygen, we studied the fluorescence of Egyptian blue nanosheets ($CaCuSi_4O_{10}$, EB-NS) in the presence of singlet oxygen. A possible effect of singlet oxygen on the NIR fluorescence of EB-NS, could indicate their suitability as nanosensors for reactive oxygen species. For the generation of $^1O_2$, we first used Rose Bengal as a photosensitizer, which efficiently generates $^1O_2$ upon irradiation with green light. This is one of the most used methods to generate $^1O_2$ and has been applied in numerous studies ranging from oxidative stress to defect chemistry in nanomaterials and photodegradation of pollutants[35,38,42]. It also offers the benefit of using the same light source that is used to excite the fluorescent nanomaterials. During these experiments, we observed spectral features expected for singlet oxygen, but also unexpected additional signals.

Upon illuminating a EB-NS dispersion with a 561 nm laser, we measured the well-known fluorescence-peak around 935 nm. Addition of Rose Bengal decreased the EB-NS fluorescence but a second spectral feature at 923 nm emerged overlapping with the EB-NS fluorescence. This feature as well as a NIR tail was also detected after addition of Rose Bengal to a control sample without EB-NS (Figure 1a).

The use of NIR fluorescent nanoparticles for plant sensing is well established, particularly for SWCNTs[15,23,31], but also for EB-NS[43]. For NIR spectra of Arabidopsis thaliana leaves (containing chlorophyll) under 561 nm laser excitation, we detected a distinct feature at 923 nm, similar to the feature observed in the Rose Bengal samples (Figure 1b, Figure S3g). This feature also appeared when pigments were extracted from the leaves into an organic solvent but was absent when spectroscopic measurements were performed on decolorized leaf tissue.
Phospholipid-based liposomes serve as an established model for biological membranes and are widely used to explore membrane interactions, permeability, and lipid organization[44,45]. We used liposomes composed of phospholipids and labeled them with the fluorescent dye nitrobenzodiazole (NBD) to study membrane behavior. Although NBD is a visible range fluorophore and not expected to produce emission in the NIR, a tail and the distinct feature at 923 nm was detected (Figure 1c). Thus, this unexpected feature occurred reproducibly despite the absence of known NIR-emitters and motivated further investigation. Spectroscopic properties of the dyes, an emission contribution of the solvents or an instrumental artifact due to the optical setup or the sample container were thus considered as possible explanations.

None of the materials used in the experiments, neither the plant tissue components nor the dyes or solvents, are known to emit fluorescence at this wavelength. Singlet oxygen appeared to be a plausible source, as it should be present in all samples including the plant leaves (photosynthesis). Interestingly, the observed wavelength of 923 nm corresponds closely to the energy gap of the second vibrational overtone of the transition from the singlet oxygen state to the ground triplet state[35]. Although this vibrational transition has previously been reported in absorption spectra, it has not, to our knowledge, been observed as an emission feature. However, the precise match in energy motivated a detailed investigation of that feature. Different possible explanations (solvents, dyes, setups) were thus tested one by one.

**Systematic investigation of the NIR feature**

In a first step, we examined whether the observed signal was specific to certain dye molecules. In addition to Rose Bengal, we tested methylene blue and tris(bipyridine)ruthenium(II) chloride (RuBPY). These three dyes possess different structures and (visible) fluorescence spectra but are all photosensitizers that generate $^1O_2$ by photoexcitation[46]. When tested under identical experimental conditions (phosphate buffered saline (PBS), excitation with a 561 nm laser), each dye showed the expected singlet oxygen phosphorescence at 1275 nm. Absolute intensities were different, which is consistent with the $^1O_2$ quantum yields reported in the literature[46]. In addition, all dyes exhibited a similar tail and feature at 923 nm (Figure 1d, Figure S1).

In a second step, we investigated different solvents to find out if solvated reactive oxygen species could be involved. In addition to PBS, we tested two polar organic solvents (methanol and acetonitrile). Again, in all cases and using the same optical setup (561 nm excitation), the 1275 nm singlet oxygen signal was detected to varying degrees depending on the solvent, and the NIR tail and the 923 nm feature was also present in each sample regardless of the solvent (Figure 1e, Figure S1). The comparison of PBS buffer based on $H_2O$ and $D_2O$ basis shows another difference. In $D_2O$ the 1275 nm peak was increased (consistent with the well-studied longer lifetime of $^1O_2$ in $D_2O$), but the 923 nm feature was reduced compared to $H_2O$ (Figure S2).

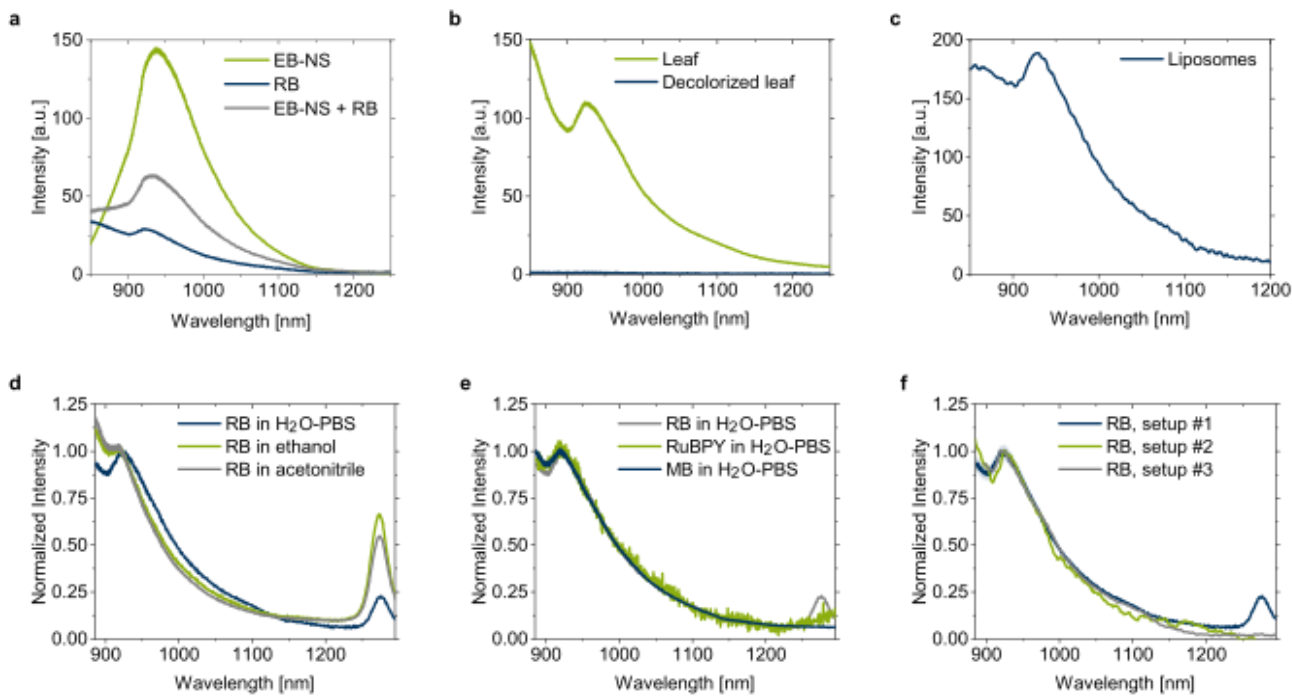

*Figure 1: Observation of a NIR emission tail and a 923 nm feature by photosensitizers/fluorophores that create singlet oxygen. **a)** Emission spectrum of Egyptian blue nanosheets (CaCuSi$_4$O$_{10}$, EB-NS) dispersed in water (0.5 mg/ml) before and after addition of Rose Bengal (RB, 100 µM), and of Rose Bengal (100 µM) in water. **b)** Emission from an arabidopsis leaf and a decolorized arabidopsis leaf (after incubation in acetone for 24 h). **c)** Emission spectrum of phospholipid-based liposomes labeled with nitrobenzodiazole (NBD). **d)** Emission spectrum of Rose Bengal in various solvents (100 µM in H$_2$O-PBS, ethanol and acetonitrile). **e)** Emission spectrum of various photosensitizers (Rose Bengal, RuBPY and Methylene Blue (MB), 100 µM in H$_2$O-PBS). **f)** Emission spectrum of Rose Bengal (100 µM in H$_2$O-PBS) measured in various optical setups with laser excitation at 561 nm (setup #1), LED excitation at 560 nm (setup #2), or white LED combined with 560±40 nm bandpass filter (setup #3). Spectra a)-e): all samples measured upon laser excitation at 561 nm.*

Finally, we tested different measurement setups to rule out the influence of specific optical components. The setups differed in excitation source (laser, LED, Hg lamp with monochromator), as well as the used optical components. InGaAs detectors were used for detection in all cases, although from different manufacturers. In addition, samples were stored and measured in both glass and polystyrene well plates. Despite these variations, the 923 nm emission feature was consistent across all configurations (Figure 1e). These tests suggested that the 923 nm emission is neither a specific property of the sensitizer molecule, nor the solvent. Furthermore, a setup-specific effect was not found.

**Detailed spectroscopic characterization**

After ruling out other causes of the signal, the emission at 923 nm continued to resemble a signal that could plausibly originate directly from singlet oxygen even though the likelihood of this optical transition should be small.

2D-excitation-emission spectra showed emission at around 923 nm as well as the known $^1O_2$ emission at 1275 nm. The feature was clearly visible, with no bands etc. that could suggest an overtone of the light source, Raman modes or similar signals as a cause (Figure 2a, Figure S3). The nature of the 923 nm emission was further investigated by calibration curves with different concentrations of Rose Bengal. This reflects the design of the original experiment to investigate how different concentrations of singlet oxygen generated by various concentrations of Rose Bengal affect the fluorescence of NIR-emitting nanoparticles. As expected, higher concentrations of Rose Bengal increased the 1275 nm $^1O_2$ emission and the 923 nm feature on top of the NIR tail increased in a proportional manner (Figure 2d, Figure S4). A similar trend was observed when the excitation laser intensity was varied. Both the $^1O_2$ emission at 1275 nm and the feature at 923 nm increased proportionally with laser intensity (Figure 2b-c, Figure S5). We then tested if the $^1O_2$ scavenger ascorbic acid can effectively reduce the concentration of $^1O_2$[38]. When ascorbic acid was added to the samples during excitation, both the characteristic $^1O_2$ phosphorescence at 1275 nm and the emission feature at 923 nm were suppressed synchronously (Figure 2a-b, Figure S6).

After dye-, solvent- and instrument-specific causes could not be found in the previous section, the similar trend under concentration and intensity variation further supported a common origin of the observed feature related to the light-induced generation of singlet oxygen.

An investigation of the temperature dependence did not provide clear results. Such a dependence would have indicated that this feature could be related to a vibrational mode. A slight increase in peak intensity was observed with increasing temperature, but this change remained close to the noise level (Figure S7).

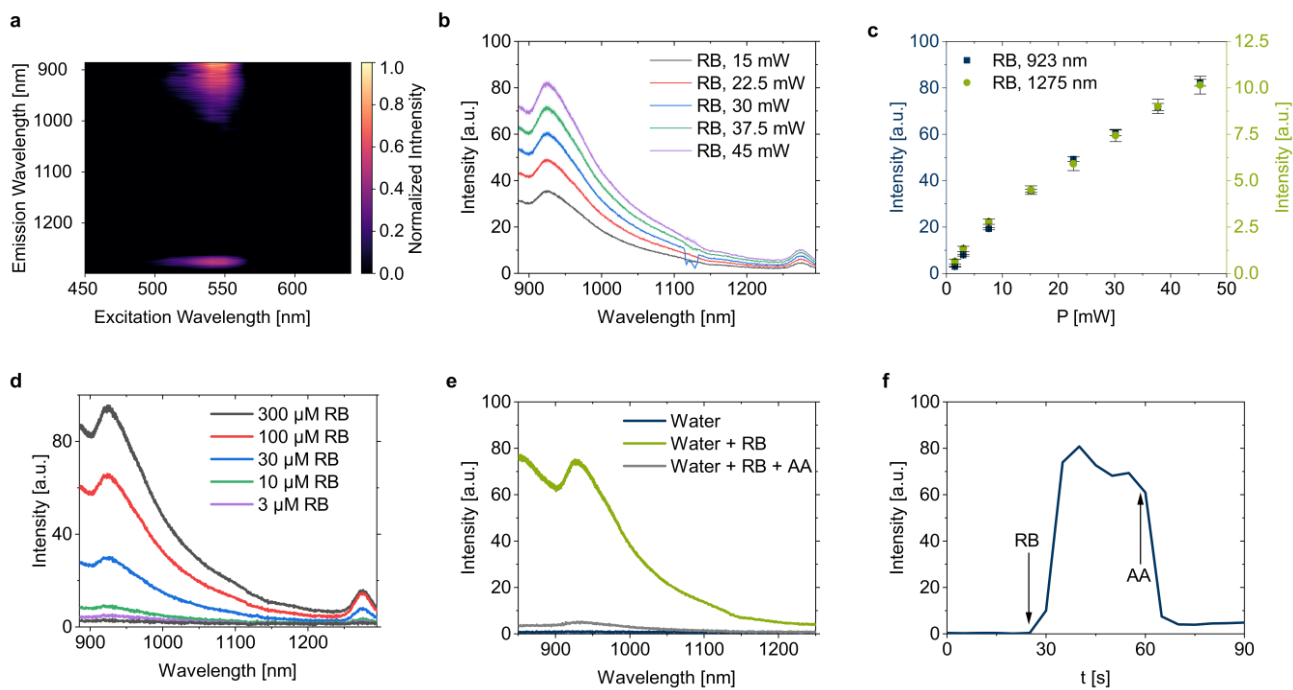

*Figure 2: Characterization of the NIR tail and 923 nm feature. a) 2D excitation-emission spectrum of Rose Bengal (RB, 300 µM in $D_2O$-PBS) upon excitation between 450 and 650 nm. b) Emission spectrum of Rose Bengal (100 µM in $H_2O$-PBS) upon laser excitation with varying intensity at 561nm. c) Intensity of spectral features at 923 nm and 1275 nm wavelength for different excitation laser intensities. Error bars = SD (n=3). d) Emission spectrum of Rose Bengal in different concentrations upon laser excitation at 561nm. e) Emission spectrum of Rose Bengal (100 µM in $H_2O$-PBS) before and after addition of ascorbic acid (AA, 10 mM) as scavenger upon laser excitation at 561nm. f) Time course for e).*

**Imaging of the NIR features**

Next, we used microscopy to image the NIR emission feature. The addition of Rose Bengal to a solution resulted in a visible increase in background intensity in a NIR fluorescence microscope (Figure 3a). In another microscopy setup, a 480 nm laser was used to excite the photosensitizer in a pattern and the NIR image of the pattern could be resolved (Figure 3b, Figure S8). Fluorescence lifetimes, for example for a RuBPY solution (Figure 3c, Figure S9), indicated a lifetime that does not fit to the longer known lifetimes of the 1275 nm singlet oxygen peak. These imaging experiments were performed with setups optimized for NIR detection up to ~1000 nm, using silicon-based detectors. Even though long-pass filtering was applied, spectral bleed-through of visible fluorescence must be considered as an option.

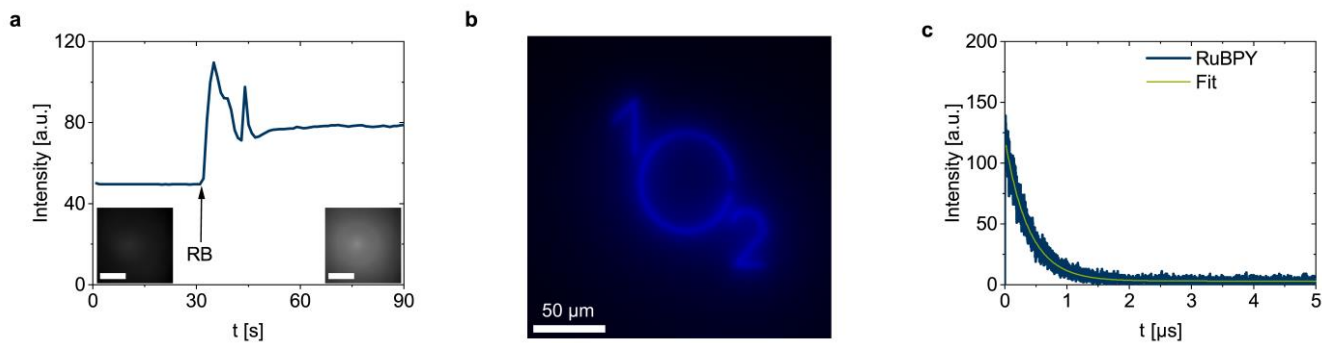

*Figure 3: Imaging of the NIR feature. a) Time course of fluorescence microscopy background intensity upon addition of Rose Bengal (RB, 300 µM) to an aqueous sample. Small images: background before and after Rose Bengal addition (scale bars = 50 µm). b) Fluorescence imaging of RuBPY solution (300 µM in $H_2O$-PBS) excited by a patterned laser beam. c) Fluorescence lifetime of a RuBPY solution (300 µM in $H_2O$-PBS) by time correlated single photon counting. Green: fitted exponential decay (τ=381 ns).*

Additionally, the measured fluorescence lifetime of 381 ns does not correspond to any known lifetime values for $^1O_2$, but matches the fluorescence lifetime of the Ru complex's visible emission[47].

**Direct or indirect correlation with singlet oxygen**

So far, the NIR tail and the 923 nm feature was observed in all experiments with photochemical generation of singlet oxygen. To further investigate its relation to $^1O_2$, we chemically generated singlet oxygen. In short, hydrogen peroxide ($H_2O_2$) was added to granulated sodium dichloroisocyanurate, resulting in the formation of chlorine gas and singlet oxygen. This reaction is characterized by its visible red luminescence associated with singlet oxygen formation.

Under these conditions, a clear emission peak was detected at 1275 nm, a visible red glow from the solution also confirmed the presence of $^1O_2$. However, unlike in the photochemical experiments, no NIR emission tail and 923 nm feature was observed (Figure 4a, Figure S10). It suggests that the NIR tail and especially the 923 nm signal is not a general feature of singlet oxygen itself, even though photochemical and chemical singlet oxygen generation pathways differ in mechanism, local environment, and excited state. Visible fluorescence spectra of the respective dyes display the known and expected emission, but no peak in the far red related to singlet oxygen (Figure 4d, Figure S11).

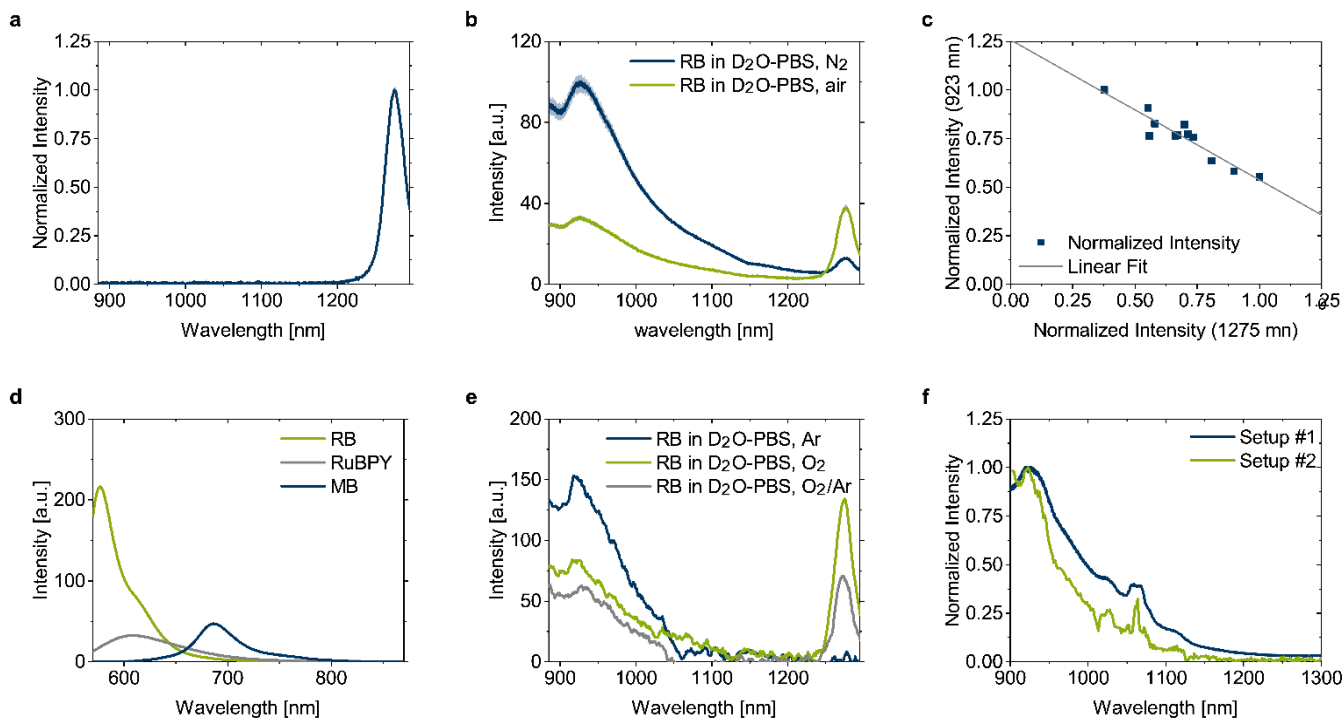

*Figure 4: Evidence for a non-singlet oxygen related feature. a) Emission spectrum of chemically generated singlet oxygen (by adding $H_2O_2$ to sodium dichloroisocyanurate). b) Emission spectrum of Rose Bengal (100µM in $D_2O$-PBS) stored under air or purged with $N_2$. c) Correlation of spectral features at 932 nm and 1275 nm wavelength for multiple samples (n=12) of Rose Bengal (100 µM in $D_2O$-PBS) purged with $N_2$ for different times. Linear fit (y=1.26-0.72x, $R^2$=0.89). d) Visible fluorescence spectra of various photosensitizers (Rose Bengal, RuBPY and Methylene Blue, 100 µM in $H_2O$-PBS). e) Emission spectrum of Rose Bengal (100 µM in $D_2O$-PBS) stored under $O_2$ or purged with Ar. f) Detected spectrum upon direct illuminating different optical setups with broad band LED. b), d): shades represent SD, n=3.*

To rule out a direct relation of the 923 nm feature to singlet oxygen, we systematically reduced the oxygen content in our samples. Nitrogen purging was performed for photosensitizer solutions in $D_2O$, to lower the dissolved oxygen concentration without removing it completely. Under nitrogen conditions, the two peaks exhibited the opposite behavior. The singlet oxygen emission at 1275 nm decreased while the 923 nm feature increased (Figure 4b-c, Figure S12). In samples containing methylene blue, which initially showed strong 923 nm emission but only weak 1275 nm singlet oxygen emission, oxygen reduction had no measurable effect. This observation suggests that the two peaks may arise from competing processes, rather than both being related to the same origin. For the Rose Bengal solution under complete deoxygenated conditions by argon purge over an extended period of time and storage under argon atmosphere the 1275 nm singlet oxygen peak disappeared entirely confirming the absence of $^1O_2$, while the 923 nm feature further increased in intensity (Figure 4eFigure S12). These results strongly support the conclusion that the 923 nm emission is not related to singlet oxygen itself but related to the fluorescence of the photosensitizers used for photochemical singlet oxygen generation. Finally, we exposed the optical setup itself to strong visible light, without any dye-containing solution. A high-intensity broadband LED was used and a NIR tail and a feature at 923 nm was detected (Figure 4f) for two different optical setups. It suggests that the 923 nm feature is not a real chemical signal even though the overall NIR tail in general correlates with singlet oxygen formation because this is correlated to fluorescence intensity.

**Mechanism and simulation of the NIR tail and 923 nm feature**

Bright visible fluorescence explains the general NIR tail across all setups. All photosensitizer dyes tested in different solvents emit strongly in the visible range, as well does chlorophyll in plant tissue under laser excitation. The concentration and laser intensity dependency correlates with the visible emission of the dyes, which correlates with the singlet oxygen produced by the sensitizers.

In contrast, no feature at 923 nm could be observed in the chemical generation of singlet oxygen where no dyes are present. Although some emission of red light was visible here, its intensity is much lower than that of the photosensitizing dyes, which explains the absence of the artifact under these conditions. Upon the absence of oxygen, singlet oxygen generation is suppressed. Thus, more energy absorbed by the dyes can be emitted as visible light. $^1O_2$ generation and visible emission indeed are competing processes when using photosensitizers.

If the oberserved feature does not originate from a vibrational mode of $^1O_2$ that is more pronounced in oxygen-deprived environment, it must be attributed to the optical setups. For Si-based imaging, insufficient filtering may be a cause. The peak shape observed in InGaAs-based spectroscopy requires a more complex explanation. Given the extremely low quantum yield of the singlet oxygen phosphorescence $(10^{-7})^{36}$ even a minor tail from visible dye emission may contribute significantly to the signal.

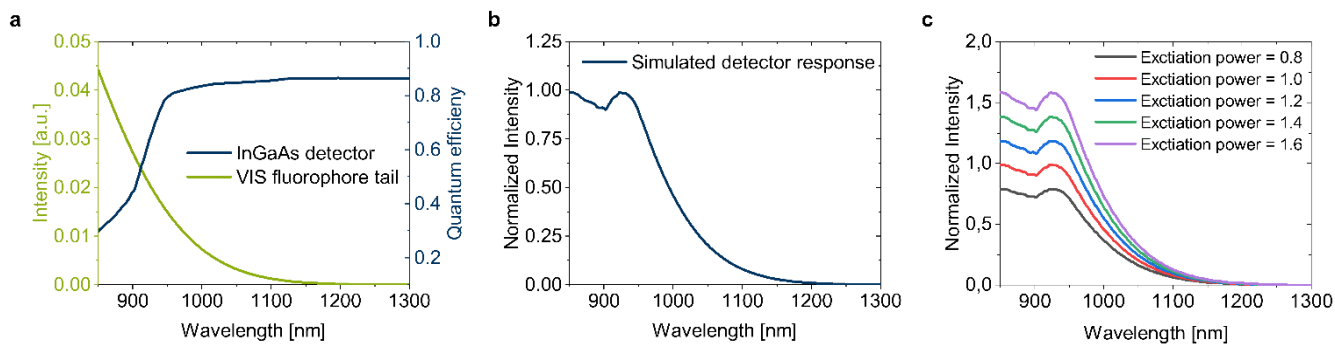

*Figure 5: NIR tail and 923 nm feature explained by a simulation. a)* NIR tail of simulated visible emission and quantum efficiency of the InGaAs detector used for NIR spectroscopy[48]. *b)* Simulated signal reported by the InGaAs detector upon strong VIS illumination, calculated by superimposing the quantum efficiency of the detector and the NIR fluorescence tail of a visible fluorophore as in a). *c)* Simulated signal measured by the InGaAs detector upon VIS illumination corresponding to increasing fluorescence intensity.

One possible explanation is the sensitivity of detectors. InGaAs detectors typically lose sensitivity below 900 nm (Figure S13). But the sensitivity shows an edge in a region around 923 nm (Figure 5a). Simulations show that an overlay of the low intensity tail of a visible emission and the sensitivity curve of InGaAs detectors leads to spectra that contains such a feature in the NIR. For the sensor used in the experiments shown in this manuscript, an apparent signal at exactly 923 nm was simulated (Figure 5b).

Another possible contribution is the optics themselves. Earlier work reported absorption of visible light and weak reemission in the NIR by the material of optical components[49]. The reflectivity of commonly used optical filters can exhibit an edge around 923 nm (Figure S14)[50]. Although it is only a small feature at low intensity, in a low signal environment this could appear as an emission feature. Although the 923 nm feature appeared across different setups and manufacturers, many optical components share similar materials suggesting a systematic origin.

While the origin of the emission at 923 nm cannot be directly attributed to singlet oxygen, there is still a strong correlation between this feature and the overall NIR tail and the generation of singlet oxygen by photosensitizers (Figure 2). This correlation only exists under well-defined constant environmental conditions, in particular with an identical solvent and a constant concentration of oxygen in the solvent. However, in many applications these constant environmental conditions are given, for example in phototherapy. In this case, a higher excitation rate of the photosensitizer leads to a higher rate of singlett oxygen generation, but at the same time also to an increase of the photosensitizers fluorescence, that can be detected via its NIR tail (Figure 5c). This could make an observation of the NIR emission tail well suited to monitor the generation of singlet oxygen by a photosensitizer. The emission of the sensitizer in the visible spectral range could as well be used, but the advantages of the NIR spectral range, such as lower background and lower scattering, outweigh the lower intentisty in the NIR. Therefore, the use of the NIR tail between 900 and 1000 nm offers a promising indirect approach to study photochemically generated singlet oxygen formation.

**Conclusion**

In summary, we observed and investigated a recurring NIR emission tail and a feature at 923 nm that initially appeared to be associated with singlet oxygen phosphorescence. Using a combination of different dyes, solvents, optical setups and experimental conditions, we systematically explored the cause of this signal. Our results indicate that the 923 nm feature is in fact not directly from singlet oxygen, but instead is due to the strong visible fluorescence of the photosensitizers/dyes that extends to the NIR. These features occur consistently under different experimental conditions and setups and may be attributed to the overlapping sensitivity of the NIR detectors or possible interactions with optical components.

These results emphasize the importance of careful validation when working in the NIR range. While the reduced background fluorescence is a key advantage of this spectral range, it can also lead to wrong conclusions at low signal intensities. Adequate controls are essential to ensure accurate data interpretation. Due to the strong correlation between the observed emission feature and the generation of singlet oxygen under constant environmental conditions, the use of this feature for monitoring singlet oxygen generation nevertheless appears to be a promising approach.

While the signal does not originate directly from singlet oxygen, it still demonstrates strong potential because of its correlation to singlet oxygen formation. The near-infrared (NIR) emission tail also decays more slowly than background and autofluorescence, making it easier to distinguish the signal in the NIR region compared to its visible peak region[51–53]. Additionally, it can be detected with standard Si detectors instead of specialized InGaAs detectors as the singlet oxygen phosphorescence.

## Methods

### NIR fluorescence spectroscopy

For the NIR fluorescence spectroscopy measurements a custom-built setup was used if not specifically stated otherwise. It consists of a Gem 561 laser (Laser Quantum), focused into a IX73 microscope (Olympus) equipped with a 20x objective. For spectroscopy, a Shamrock 193i spectrograph (Andor Technology) coupled to an array NIR detector (Andor iDUs InGaAs 491) was connected to the microscope. If not stated otherwise, for spectroscopy measurements, 200 µl of the respective solutions were placed in a 96-well plate and positioned above the objective. Fluorescence data was acquired via the Andor SOLIS software. The laser power was set to 100 mW, exposure time to 10 s and input slit width to 500 nm. Data analysis and plotting was performed using Origin Lab. The excitation of 2D excitation–emission spectra was performed with a monochromator (MSH 150, Quantum Design GmbH) equipped with an LSE341 light source (LOT-Quantum Design GmbH) and recorded with the same setup described above.

The second setup, that was used for control measurements, is another custom-made setup illuminated by a 565 nm LED (Thorlabs M565L3), focused by a 20x objective (Olympus LUCPlanFL N, 20x) on a well-plate holder. The NIR signal collected by the objective is separated by a dichroic mirror (Thorlabs DMLP805) and fiber-coupled into a NIR spectrometer (Ocean Insight, NIRQUEST+1.7).

For another control experiment, a third setup was used. It was performed at an inverted microscope (Nikon Eclipse Ti2). It consisted of an excitation light (CoolLED pE300 Lite, 100% power), which was transmitted through a 560 nm bandpass filter(AHF Analysentechnik F47-561) and an 804 nm longpass dichroic mirror (AHF Analysentechnik F38-801) before being focused by a 20x objective (CFI S P-Fluor ELWD ADM, Nikon). Emission light passed through an 840 nm longpass filter (AHF Analysentechnik F47-841) and was coupled into a NIR spectrometer ( AVASPEC-NIR256-1.7-HSC-EVO, Avantes)

### NIR fluorescence imaging

For NIR fluorescence microscopy, a custom-built set-up was used. An inverted microscope (Olympus IX73) equipped with a 20x objective, coupled to a 561 nm laser (Cobolt Jive™) for excitation was used. NIR fluorescence images were filtered with two 900 nm long pas filters and recorded with an Si camera (Excelitas PCO edge 4.2 bi).

**Patterned NIR fluorescence imaging**

Microscopy was performed using a Thunder Imager DMi8 (Leica Microsystems, Germany). Excitation was performed at 480 nm laser quipped for patterning selected areas of the sample. A 63x oil immersion objective (HC PL APO 63x/1,40-0,60 OIL, Leica, Germany) was used. The images were obtained from Leica LAS X Thunder software and analyzed with ImageJ/Fiji.

**Fluorescence lifetime measurements**

Fluorescence lifetime measurements were performed in a MicroTime 200 system (PicoQuant, Berlin, Germany) consisting of a pulsed laser (641 nm, PicoQuant) an Olympus IX73 inverted confocal laser scanning microscope equipped with an 60x water-immersed IR objective (Olympus) and SPAD (Excelitas Technologies, Mississauga, Kanada) detectors. The samples were excited with the pulsed laser (650 nm, 1 MHz). The emitted light was separated by a dichroic mirror (R405/488/532/635, Semrock, Illinois, USA) from the excitation light, passed a 900 nm longpass filter (FEL0900, Thorlabs), and was focused on a SPAD detector. Time correlated single photon counting histograms were recorded using a HydraHarp400 (PicoQuant) system. TCSPC data were fitted with a exponential decay function and considering the corresponding instrument response function.

**Plant grows and leaf pigment extraction**

*Arabidopsis thaliana* seeds were sown in 8×8 cm pots filled with pre-mixed soil (BioBizz All-Mix). After germination, one seedling was retained per pot, and plants aged 4–5 weeks were used in this study. The plants were grown on a custom-built growth rack with approximately 30 cm spacing between each shelf, each equipped with five LED light tubes (Bioledex GoLeaf LED Plant Light). Fresh leaf tissue was chopped and placed into 5 ml Eppendorf tubes, followed by the addition of 3 ml of acetone. The tubes were shaken for 24 h in the dark on a shaker to extract chlorophyll *in vitro* and obtain decolorized leaf.

**Acknowledgements**

Funded by the Deutsche Forschungsgemeinschaft (DFG, German Research Foundation) under Germany´s Excellence Strategy – EXC 2033 – 390677874 – RESOLV. This work is supported by the "Center for Solvation Science ZEMOS" funded by the German Federal Ministry of Education and Research BMBF and by the Ministry of Culture and Research of Nord Rhine-Westphalia. We thank the DFG for funding. Moreover, we want to thank Prof. Peter R. Ogilby and Dr. Mikkel Bregnhøj from Aarhaus University for insightful discussions in singlet oxygen.

**Supplementary Information**

*A near infrared emission feature from visible fluorescence tails and its correlation with singlet oxygen*

*Bjoern F. Hill[1], Jiaqi Li[1], Norman Labedzki[1], Christina Derichsweiler[1,2], Janus A. C. Wartmann[1], Krisztian Neutsch[1], Luise Erpenbeck[3], Sebastian Kruss,[1,2] ***

[1] Department of Chemistry and Biochemistry, Ruhr Universität Bochum, 44801 Bochum, Germany

[2] Fraunhofer Institute for Microelectronic Circuits and Systems, 47057 Duisburg, Germany

[3] Department of Dermatology, University Hospital Münster, 48149 Münster, Germany.

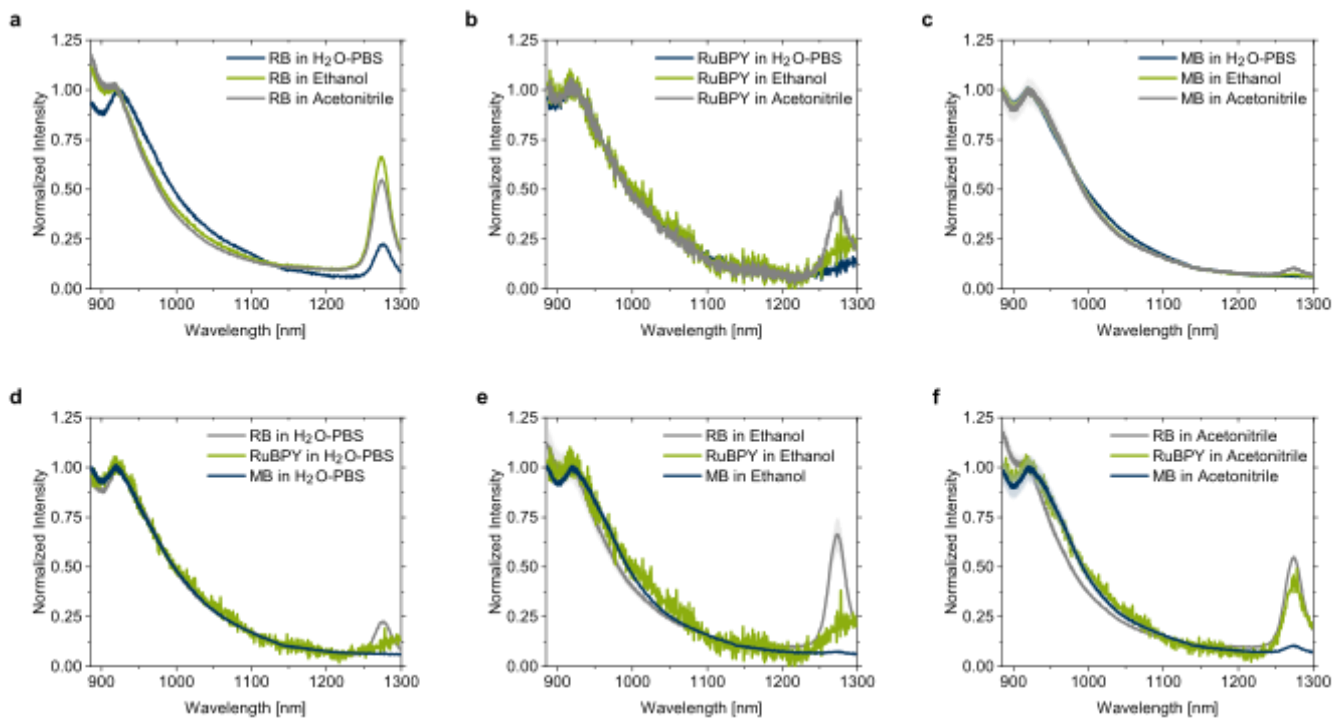

**Figure S1: Dye and Solvent screening**
*a)* Emission spectrum of Rose Bengal in various solvents (100μM in H$_2$O-PBS, Ethanol and Acetonitrile) upon laser excitation at 561 nm. *b)* Emission spectrum of RuBPY in various solvents (100μM in H$_2$O-PBS, Ethanol and Acetonitrile) upon laser excitation at 561 nm. *c)* Emission spectrum of Methylene Blue in various solvents (100μM in H$_2$O-PBS, Ethanol and Acetonitrile) upon laser excitation at 561 nm. *d)* Emission spectrum of various photosensitizers (Rose Bengal, RuBPY and Methylene Blue, 100 μM in H$_2$O-PBS) upon laser excitation at 561nm. *e)* Emission spectrum of various photosensitizers (Rose Bengal, RuBPY and Methylene Blue, 100 μM in Ethanol) upon laser excitation at 561 nm. *f)* Emission spectrum of various photosensitizers (Rose Bengal, RuBPY and Methylene Blue, 100 μM in Acetonitrile) upon laser excitation at 561 nm. All spectra: shades represent error (SD, n=3)

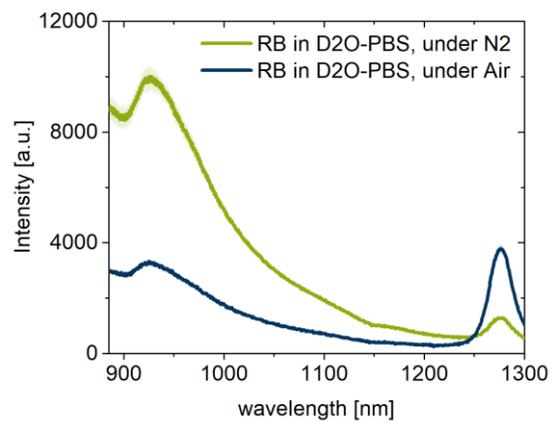

**Figure S2: Dye and Solvent screening**
*Emission spectrum of Rose Bengal in buffer solutions (100 µM in $H_2O$-PBS and $D_2O$-PBS) upon laser excitation at 561nm. All spectra: shades represent error (SD, n=3)*

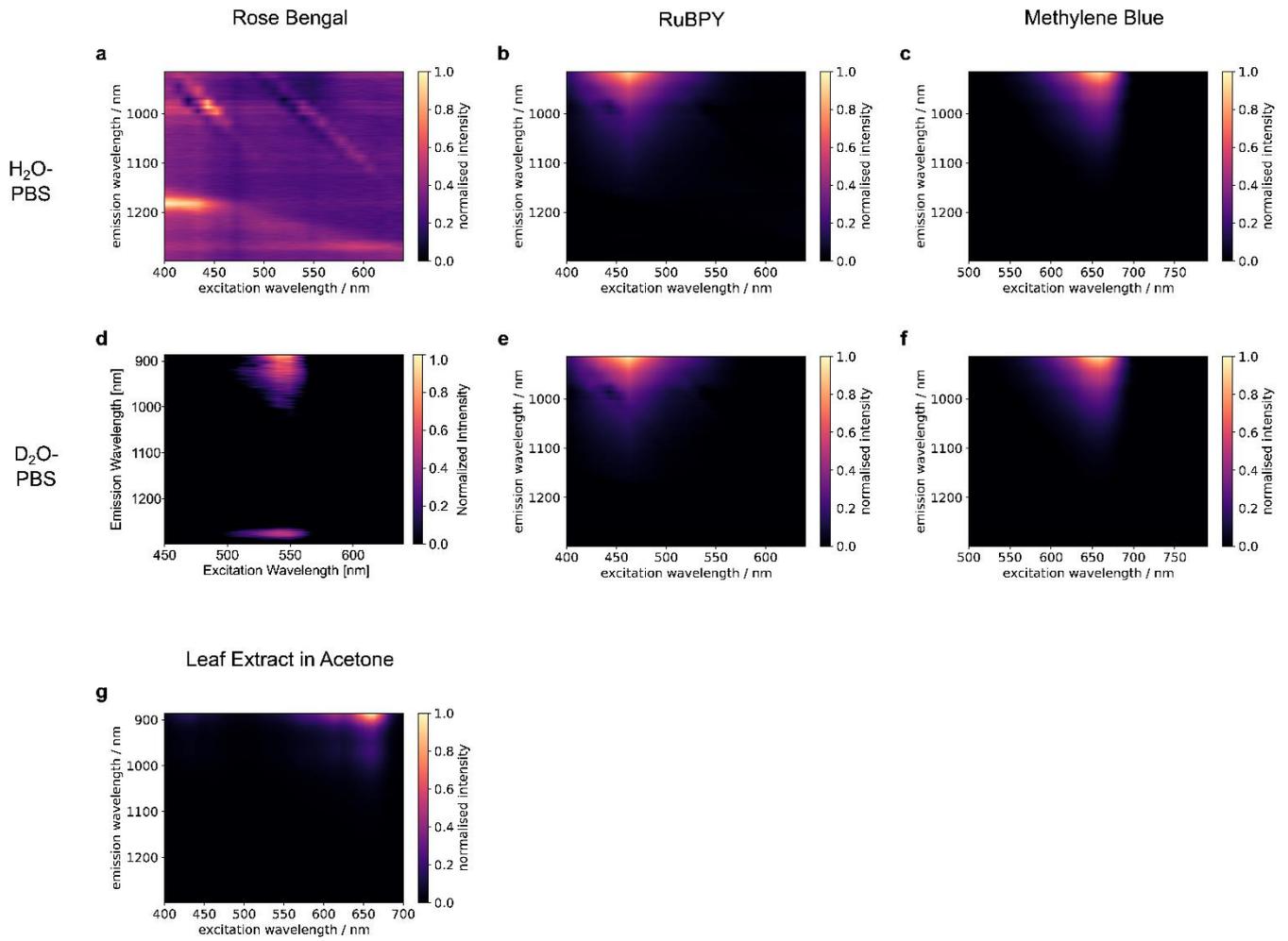

**Figure S3: 2D-spectra**
*2D Excitation-emission spectrum of various photosensitizers (a,d: Rose Bengal; b,e: RuBPY; c,f: Methylene Blue) in buffer solutions (a-c: 300 µM in $H_2O$-PBS, d-f: 300 µM $D_2O$-PBS), and of plant constituents extracted from arabidopsis leaf (by incubation in acetone for 24 h).*

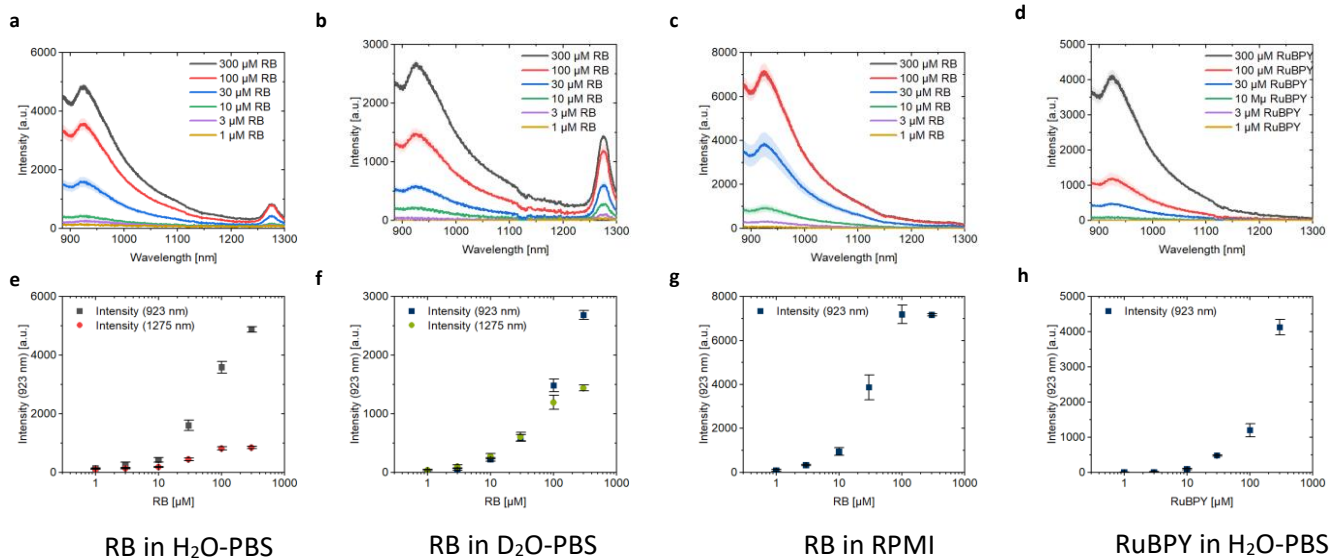

**Figure S4: Calibration Curves**
*a)-d): Emission of Rose Bengal in $H_2O$-PBS (a), Rose Bengal in $D_2O$-PBS (b), Rose Bengal in RPMI (c), RuBPY in $H_2O$-PBS (d) in different concentrations upon laser excitation at 561nm. e)-h): Intensity of spectral features at 932 nm and 1275 nm wavelength for different concentrations of Rose Bengal, corresponding to a)-d).*

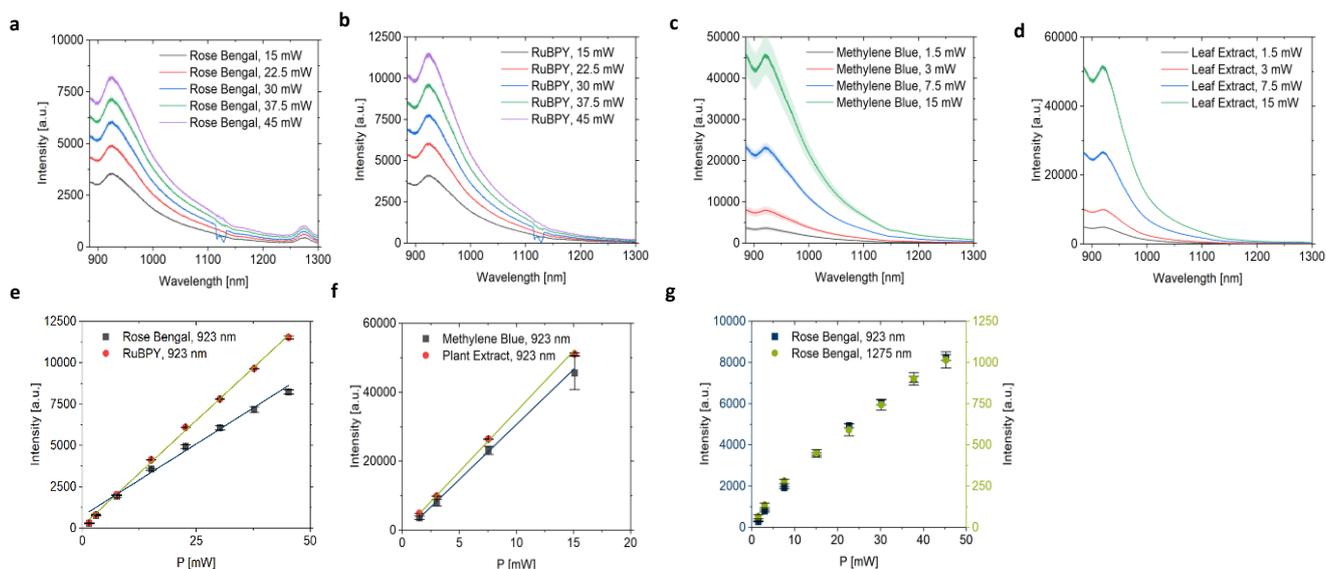

**Figure S5: Laser Intensity Curves**
*a) Emission of Rose Bengal (100 µM in $H_2O$-PBS) upon laser excitation of different intensities at 561 nm. b) Emission of RuBPY(100 µM in $H_2O$-PBS) upon laser excitation of different intensities at 561nm. c) Emission of Methylene Blue (100 µM in $H_2O$-PBS) upon laser excitation of different intensities at 561nm. d) Plant constituents extracted from arabidopsis leaf in acetone upon laser excitation of different intensities at 561nm. e) Emission Peak at 923nm of Rose Bengal and RuBPY (100 µM in $H_2O$-PBS) upon laser excitation of different intensities at 561nm. f) Emission Peak at 923nm of Methylene Blue (100 µM in $H_2O$-PBS) and plant constituents extracted from arabidopsis leaf in acetone upon laser excitation of different intensities at 561nm. g) Emission Peaks at 923 nm and 1275 nm of Rose Bengal (100 µM in $H_2O$-PBS) upon laser excitation of different intensities at 561nm.*

**Figure S6: Scavenging**
*a) Emission spectrum of RuBPY (100 µM in $H_2O$-PBS) before and after addition of Ascorbic acid (10 mM) as scavenger upon laser excitation at 561nm. b) time course for e). All spectra: shades represent error (SD, n=3)*

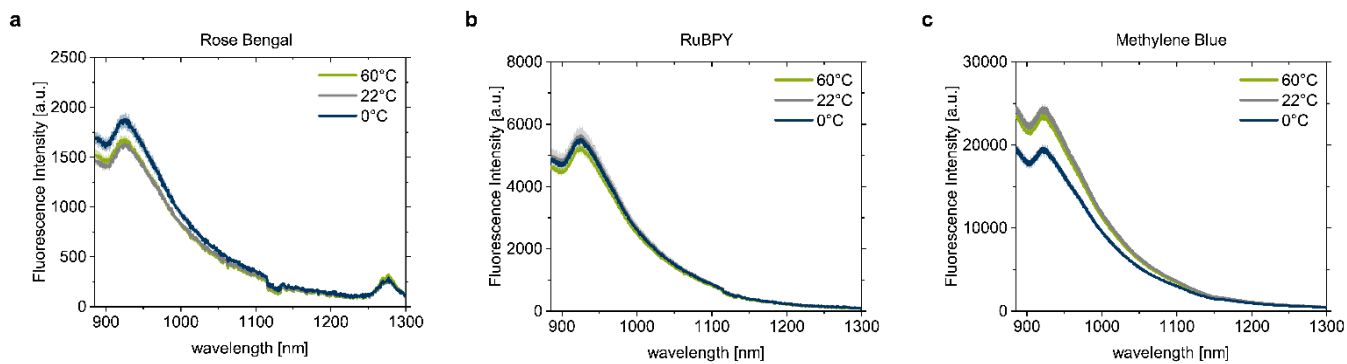

**Figure S7: Temperature Dependance**
*a) Emission of Methylene Blue (100 µM in H$_2$O-PBS) at different temperatures upon laser excitation at 561 nm. b) Emission of RuBPY (100 µM in H$_2$O-PBS) at different temperatures upon laser excitation at 561 nm. c) Emission of Methylene Blue (100 µM in H$_2$O-PBS) at different temperatures upon laser excitation at 561 nm.*

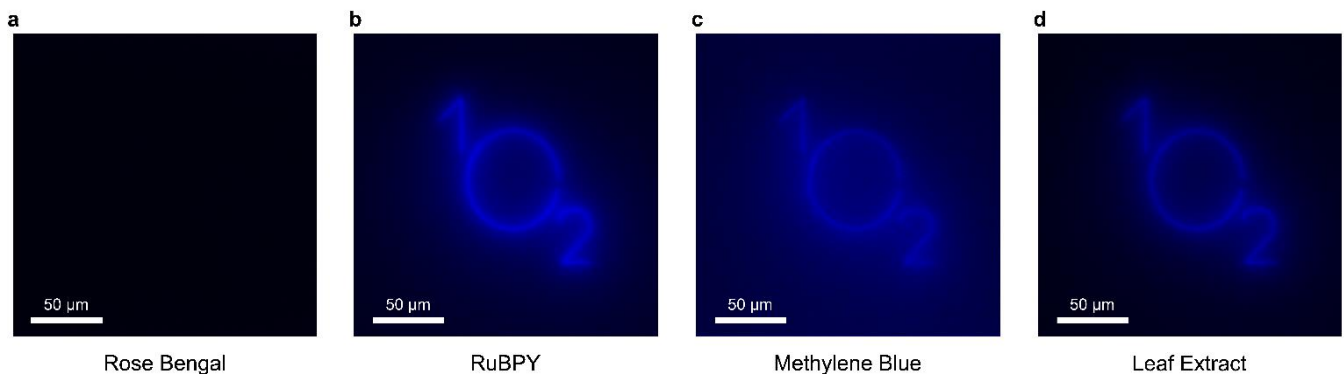

**Figure S8: Laser Patterning**
*a) Fluorescence Imaging of Rose Bengal solution (300 μM in $H_2O$-PBS) excited by patterned laser. b) Fluorescence Imaging of RuBPY solution (300 μM in $H_2O$-PBS) excited by patterned laser. c) Fluorescence Imaging of Methylene Blue solution (300 μM in $H_2O$-PBS) excited by patterned laser. d) Fluorescence Imaging of plant constituents extracted from arabidopsis leaf (by incubation in acetone for 24 h) excited by patterned laser.*

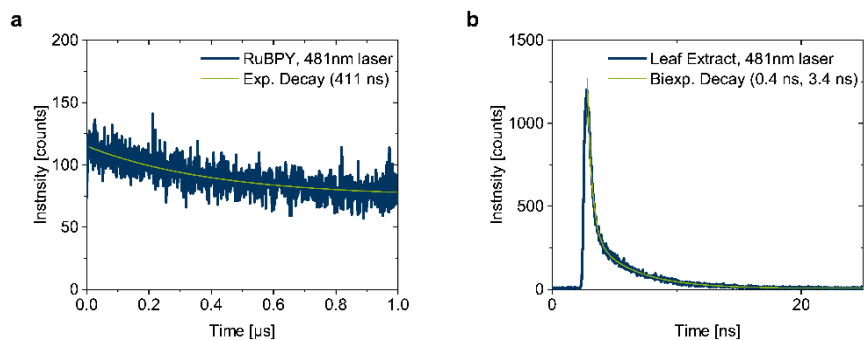

**Figure S9: TCSPS/Fluorescence Lifetime**
*a) Fluorescence lifetime of RuBPY (100 µM in $H_2O$-PBS) measured by TCSPC upon pulsed (1 MHz) laser excitation at 483 nm. b) Fluorescence lifetime of plant constituents extracted from arabidopsis leaf (by incubation in acetone for 24 h) measured by TCSPC upon pulsed (40 MHz) laser excitation at 483 nm.*

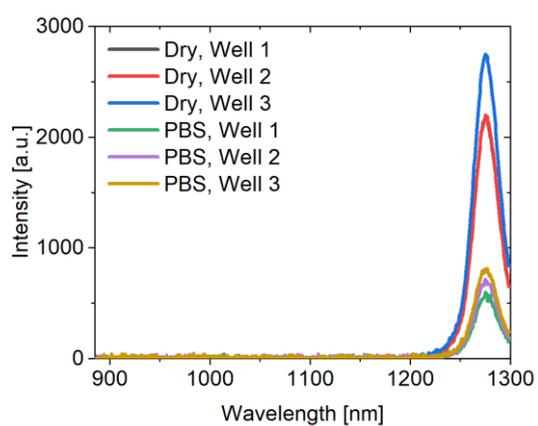

**Figure S10: Chemical Generation of Singlett Oxygen**
*Emission of chemically generated singlett oxygen under different condtions: By adding $H_2O_2$ to granulated sodium dichlorisocynurate („dry") or to sodium dichlorisocynurate dispersed in $H_2O$-PBS („PBS")*

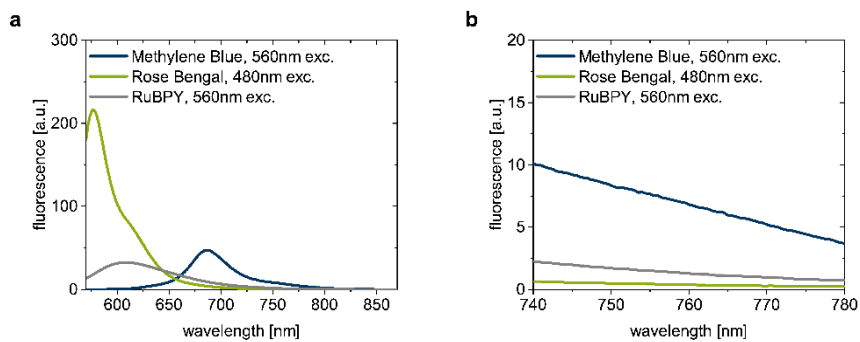

**Figure S11: VIS fluorescence spectra**
*a) Visible Fluorescence spectra of various photosensitizers (Rose Bengal, RuBPY and Methylene Blue, 100 µM in $H_2O$-PBS). b) Detail of a) around the $^1O_2$ excitation wavelength of 760 nm.*

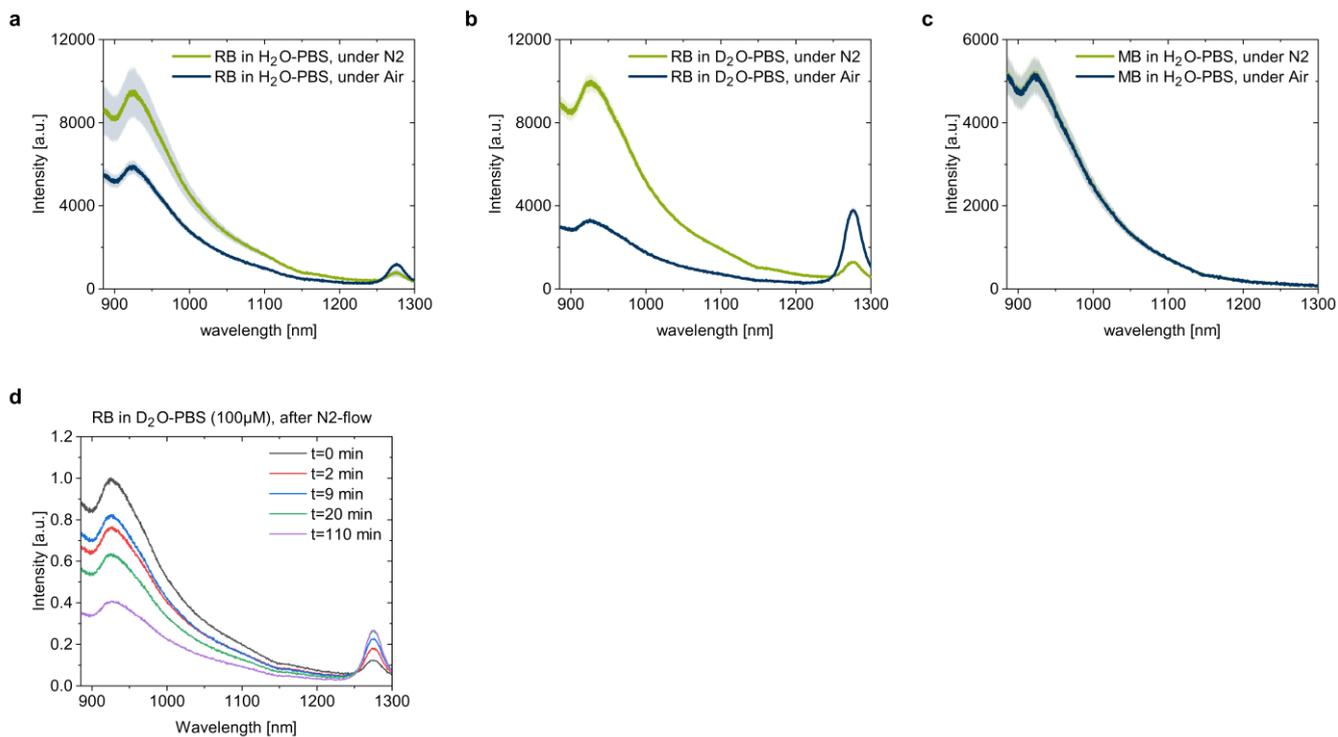

**Figure S12: Nitrogen purge**

*a) Emission spectrum of Rose Bengal (100 µM in $H_2O$-PBS) stored under air or purged with $N_2$. b) Emission spectrum of Rose Bengal (100 µM in $D_2O$-PBS) stored under air or purged with $N_2$. c) Emission spectrum of Methylene Blue (100 µM in $H_2O$-PBS) stored under air or purged with $N_2$. d) Emission spectra of Rose Bengal (100 µM in $D_2O$-PBS) stored under Air for different times after purged with $N_2$.*

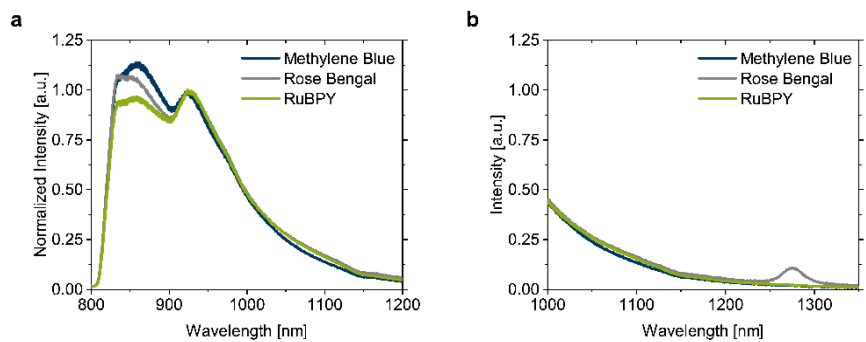

**Figure S13: Full detector range**
*a),b) Emission spectrum of various photosensitizers (Rose Bengal, RuBPY and Methylene Blue, 100 µM in $H_2O$-PBS) upon laser excitation at 561 nm.*

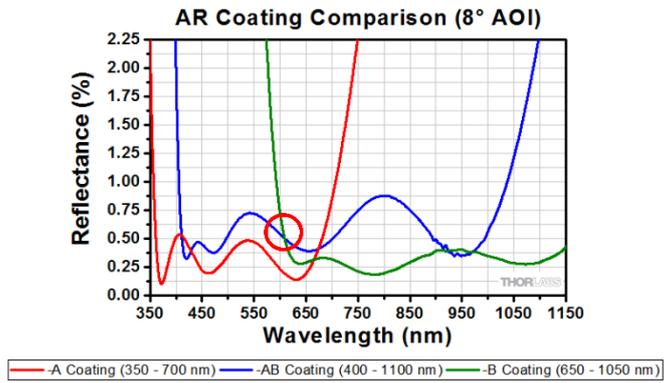

**Figure S14: Optical properties of standard lenses**
*Reflectance of commercially coated lenses used in IR-spectroscopy setups[48]*